# The Nonlinear Dynamic Conversion of Analog Signals into Excitation Patterns

Gerold Baier and Markus Müller

*Facultad de Ciencias, Universidad Autonoma del Estado de Morelos, 62210 Cuernavaca, Morelos, Mexico*

*e-mail:* baier@servm.fc.uaem.mx

## Abstract

Local periodic perturbations induce frequency-dependent propagation waves in an excitable spatio-temporally chaotic system. We show how segments of noise-contaminated and chaotic perturbations induce characteristic sequences of excitations in the model system. Using a set of "tuned" excitable systems, it is possible to characterize signals by their spectral composition of excitation pattern. As an example we analyze an epileptic "spike-and-wave" time series.

PACS numbers: 05.45.Gg; 05.45.Tp; 05.45Xt

For systems of diffusively coupled nonlinear oscillators it was found that excitable spatio-temporal chaos may undergo a generic pattern transition when perturbed locally at an appropriate frequency [1]. The switching from a globally disordered pattern of low mean amplitude to a globally ordered pattern of traveling excitation waves can be used for the "detection" of the perturbation frequency. To do this one can simply record the induced excitations, i.e. the suprathreshold firing of one of the model's elements.

A related detection principle is exploited in the time series analysis performed by the ear (see e.g. [2]). In the mammalian inner ear, for example, sound-induced mechanical vibrations are converted into sequences of nerve cell firings that propagate as excitation waves to the auditory cortex. There, the temporal patterns of excitations are interpreted with high accuracy and noise-tolerance, in spite of the original sound signal's complexity and nonstationarity. The common assumption in the modeling of this process is that the spikes are induced from a resting state that is an excitable fixed point [3]. However, it was shown recently that both spontaneous and induced neural excitation patterns can be modeled with a resting state that is low-dimensional chaos [4]. Furthermore, it was shown that this might even *improve* the reliability of induced firing.

We have suggested that the pattern transition reported in [1] be used as a device to analyze time series that are noise-contaminated and nonstationary [5]. So far, the following problems have not been dealt with: i) How does the response of such a system depend on perturbation amplitude and frequency; ii) how does the system react in the presence of noise in the signal; iii) what is the temporal resolution, and iv) how is the response to artificial non-harmonic (e.g. deterministic chaotic) and experimental perturbations. Here, we study these questions using a prototypic model system.

The model is a set of nonlinear excitable oscillators coupled by (linear) diffusion in one spatial dimension in the following form:

$$\frac{1}{\varepsilon}\frac{dX_1}{dt} = f(X_1, Y_1) + D_X(X_2 - X_1) + pert/(A\varepsilon)$$
$$\frac{1}{\varepsilon}\frac{dY_1}{dt} = g(X_1, Y_1)$$

$$\frac{1}{\varepsilon}\frac{dX_i}{dt} = f(X_i, Y_i) + D_X(X_{i+1} + X_{i-1} - 2X_i)$$
$$\frac{1}{\varepsilon}\frac{dY_i}{dt} = g(X_i, Y_i) \qquad (1)$$

$$\frac{1}{\varepsilon}\frac{dX_N}{dt} = f(X_N, Y_N) + D_X(X_{N-1} - X_N)$$
$$\frac{1}{\varepsilon}\frac{dY_N}{dt} = g(X_N, Y_N)$$

for $i = 2, 3, ... N-1$, where $N$, the number of oscillators, is chosen to be 30, and the boundary conditions are zero-flux. Parameter $\varepsilon$ is a velocity constant that is identical for all variables of the unperturbed differential equation. For a given external perturbation the model velocity (and thereby the model's internal frequencies) can be adjusted by changes in $\varepsilon$. The additive term *pert* (not affected by changes of $\varepsilon$) denotes the local external perturbation. Its strength is controlled by coupling parameter $A$. The nonlinear functions *f* and *g* are adapted from the Goldbeter-Dupont-Berridge model [6]:

$$f(X,Y) = a - m_2 X/(1+X) + m_3 YX^2/((k_1+Y)(k_a+X^2)) + Y - X$$
$$g(X,Y) = m_2 X/(1+X) - m_3 YX^2/((k_1+Y)(k_a+X^2)) - Y$$

In the absence of external perturbation the chain of oscillators has spatio-temporally hyperchaotic solutions [1]. With model parameters fixed in the Canard region of the individual oscillator [7] the chaos is *excitable*, i.e. a single short perturbation exceeding some threshold value leads to a spike of significantly larger amplitude than the spontaneous oscillations (see [8]). However, in spite of being an excitable medium the exponential divergence of the basal chaos prevents any long-range propagation of waves following a single local suprathreshold perturbation.

If a sinusoidal periodic perturbation is added to the equation for variable $X_1$ of the first oscillator, *pert*=sin($\omega$t) with amplitude $1/A$, excitation waves can be induced for some frequency windows of the perturbation [1]. These large-amplitude waves propagate from the point of perturbation to the far end of the chain. Fig. 1 is a bifurcation diagram showing maxima of variables $X_1$ and $X_{30}$ as a function of forcing frequency $\omega$. In Fig. 1a the two dark regions in the plot of $X_1$ (1.5<$\omega$<2.3, and 3.1<$\omega$<4.5) indicate perturbed (hyper-)chaotic behavior with a distribution of maxima that is wider than in the chaotic resting state. In contrast, the central region (2.3<$\omega$<3.1) indicates periodic or quasiperiodic behavior of medium amplitude. It is in this window that excitation waves are generated. In Fig. 4b the left dark region and the far right region in the plot of $X_{30}$ (1.5<$\omega$<2.3 and 4.0<$\omega$<4.5, respectively) are due to chaotic behavior with a distribution of maxima that is equal to the chaotic resting state. In these bands the influence of the perturbation is much weaker than in Fig. 1a. The region 3.1<$\omega$<4.0 is a mixture of basal chaos and irregularly induced excitation waves. The central region (2.3<$\omega$<3.1) of Fig. 1b is formed by periodic oscillations of large amplitude. In this window the induced large-amplitude excitation waves reach the far end of the chain of oscillators. The periodic window maintains its size in the frequency domain throughout the chain. The small region of complex bifurcations within the periodic window in the center of Fig. 1a is lost in the course of propagation. From oscillator 8 onward the propagation waves are fully established and from oscillator 16 onward the bifurcation structure remains qualitatively unchanged. The change in bifurcation structure as a function of the number of the individual oscillator thus shows how the spatio-temporal arrangement of excitable units acts as a kind of sharp-edged band-pass filter.

For *A>95* no excitation waves are detected in oscillator 30, i.e. a minimum coupling strength is required to start the process of propagation. For *80<A<95* the frequency window of induced excitations widens as coupling strength is increased. For 40<A<80 the frequency window remains nearly constant in width. In addition, for stronger couplings new windows with more complex excitation waves, like period 2 and period 3, appear.

The periodic excitation waves do not stabilize an unstable periodic orbit of the unperturbed system [1]. Therefore the induced excitation waves do not follow an orbit of the unperturbed system and are not an example of successful chaos control by external periodic forcing as introduced by Pyragas [9].

The Fourier spectrum of the unperturbed chaos in eq. (1) has a broadband distribution with well-pronounced maxima at about 3.4 and 3.6 s$^{-1}$. The frequency range for which induction of excitation waves is found does not coincide with these maximum power frequencies. If equation (1) is scanned with a coupling of the sine wave for which no periodic excitation waves are induced at all, the mean amplitude of the response does not increase in the range $2.3<\omega<3.1$. If Fourier spectra are calculated for this case we observe a clear maximum of the power in the frequency range where excitation waves are induced for stronger couplings. However, the power maximum is not due to an increase in amplitude of the respective Fourier component (as in classical resonance). It is a consequence of the prolonged fraction of time that the system spends near the "ghost" of the periodic solution and thus indicative of the nearby crisis. These results thus differ from classical resonant firing as seen e.g. in the periodically forced Hodgkin-Huxley equation [10].

Analysis of the bifurcations that lead to excitation waves shows that both at $\omega\approx2.3$ and $\omega\approx3.1$ the chaotic attractor undergoes a crisis. The crises occur without multistability. Both the period 1 and period 2 attractors are created at the critical frequencies in a saddle-node bifurcation of periodic cycle. If the perturbation frequency is chosen within the excitation window and initial conditions are selected on the unperturbed chaotic attractor, chaotic transients are observed that follow trajectories similar to the unperturbed case for some time. Within this frequency region a chaotic attractor that closely resembles the attractor seen e.g. at $\omega=2.2$ can be stabilized if the unperturbed chaotic signal is used as external driving. The mechanism that locally suppresses the chaos thus requires that the periodic forcing destabilizes the chaotic solution in a crisis and turns it into a chaotic saddle.

All solutions of the first oscillator found within the frequency window where the chaotic solution is a saddle, have a dominant Fourier component at the frequency of the driving sine wave. The resulting non-chaotic solution then serves as a driving with this frequency to its neighbor in the chain of oscillators. The correct choice of coupling constant between oscillators then succeeds to turn the chaos of the neighboring oscillator into a saddle yielding another element with dominant frequency of the sine wave. Thereby the process of propagation is initiated and continues to the end of the chain if all oscillators are identical.

Induction of periodic waves in gradient-free spatio-temporal chaos by means of local periodic perturbation was found in the case of the complex Ginzburg-Landau equation [11] and in models of CO oxidation and of cardiac tissue activity [12]. In the former case, periodic forcing in an area around the spiral tip stabilized an unstable spiral. In the latter case, target waves were created for a finite frequency window around the maximum power of the unperturbed chaos and a mechanism of nonlinear resonance was postulated. After switch-off of the control the model systems evolved to a homogeneous fixed point, which is in contrast to the present case where the spatio-temporal chaos is the only attractor of the unperturbed eq. (1).

Now we study the dependence of the induction of excitation waves in the presence of noise in the perturbation. The chosen perturbation consists of Gaussian white noise (mean zero and amplitude variance equal to 1) to which a sequence of 30 full cycles of the sine wave of amplitude 1 is added. Fig. 2 shows scans of two parameter planes for eq. (1) with this perturbation. In order to obtain results that are independent of the particular choice of initial conditions the average of 30 runs with different initial conditions is evaluated for each set of parameters. This way, the effect of randomly induced excitation waves (especially near the high-frequency end of the periodic window in Fig. 1b) is suppressed. Successful induction is characterized by the number of induced excitation waves exceeding a threshold value in oscillator 30. The threshold is set as 0.7 of variable $X_{30}$ (a value that is not exceeded in the averaged chaotic basal state).

In the ω/ε-plane (Fig. 2a) a single band is found where excitation waves are induced whereas the rest shows no excitation at all. The band is displaced linearly and widens slightly as the frequency is increased. The relationship between velocity factor and perturbation frequency at maximum induction is given in linear regression as ε=2.6ω. The horizontal width of the band allows an estimation of the frequency resolution. The resolution is better than in the case of a pure sine wave (Fig. 1) because the noise tends to impede the induction of excitation waves and this effect is stronger near the borders of the induction band. Furthermore, the results in Fig. 2a are averages over 30 runs with different initial condition. This averaging is essential to improve the frequency resolution compared to Fig. 1b. The chaotic excitations adjacent to the region of regular excitation are thereby suppressed. Fig. 2b shows the scan of parameter plane *A*/ω at fixed ε. Here, parameter *A* controls the amplitude of the combined noise/sine wave perturbation and does not affect the relative amplitude of the two contributions. In Fig. 2b, the dark tongue centered at ω≈2.8 Hz represents a unique induction zone for couplings *A*>40. For stronger couplings (*A*<40) the structure gets more complex as harmonics of the optimal perturbation frequency also succeed in inducing waves. The first harmonic is visible as a light gray zone at ω≈5.6. At even lower values of *A* new bifurcations lead to more complex types of excitation waves, e.g. period 2 and period 3 solutions as in the case of pure sinusoidal perturbation (see [13]).

The unperturbed system is not a classical excitable medium. Its basal state is spatio-temporally chaotic, i.e. it displays an irregular temporal evolution in all variables. We perturbed eq. (1) with the combined sine wave/white noise signal as in Fig. 2 and calculated how long it takes (averaged over 30 runs) for the first induced excitation wave to reach oscillator 30 after the onset of the sine wave. Fig. 3a shows that the average propagation time depends on forcing frequency like $1/\omega$. Plotting the result in terms of number of sine wave cycles in the perturbation a constant of 11-12 cycles is found. On the average this number of perturbation cycles in oscillator 1 are completed before the first excitation wave reaches oscillator 30. (The resulting time delay $\Delta t$ is used to shift the time axes for each frequency in Figs. 4 and 5 accordingly.)

So far, the signal to be detected was a sinusoidal oscillation. Strictly periodic and harmonic signals do not represent adequately the wide variety of natural perturbations, however. A second, more complex class of signals is represented by deterministic chaos. Chaos displays irregular variations in amplitude and frequency of individual oscillations. Due to its deterministic origin chaos may contain larger amounts of information than periodic signals, for instance in the form of near-periodic "motifs", characteristic sequences of only a few oscillations determined by the presence of weakly unstable periodic orbits within the chaotic attractor. We are therefore interested to find out whether eq. 1 is able to deal with the irregularity of a chaotic signal.

As the simplest representative of deterministic chaos we applied a time series from the Rössler equation as a perturbation. The time series was normalized to zero mean and variance equal to 1 in order to have a well-defined scale for the choice of the coupling constant *A*. As we are now dealing with a broader distribution of frequencies, the same perturbation is applied in parallel to a set of "tuned" equations (1). The tuning consists in a stepwise increase of velocity constant $\varepsilon$ in steps of 0.05. Therefore each chain of oscillators has its own preferred response frequency and the whole set covers a frequency band from 0.8 to 4.2 Hz.

Fig. 4 is the excitation response in this band as a function of time. Each bar in the figure represents the times when an (averaged) excitation wave in oscillator 30 exceeds the threshold. The chaotic time series was applied during the whole period. It induces strongly between 2.6 and 3.3 Hz yielding a broad textured excitation band. The center of this band at about 3 Hz coincides with the dominant Fourier frequency of the chaotic motion. A second, thinner frequency band at about 1.5 Hz can be assigned to a subharmonic of this dominant frequency. This is consistent with the observation of oscillations that are transiently close to the unstable period 2 solution of the chaos. The different widths of the bands reflect the unequal contributions of the two frequencies to the composite signal. We find that the maximal value of coupling constant $A$ where excitation is induced is much higher for the 3 Hz component. This possibly reflects its larger power (compared to that of the subharmonic) in the Fourier transformation.

The response of subthreshold spatiotemporal chaos to local chaotic perturbations has not been studied previously. Apart from the fact that it is possible to characterize the main frequencies of a stationary chaotic signal, this is significant because the noise tolerance (Fig. 1) and the temporal resolution (Fig. 2) of the present system should allow the detection a transient chaos in experimental signals. To study this, a real world recording is applied as a perturbation.

We chose an epileptic oscillation as recorded by scalp EEG. This oscillation has been analyzed with respect to its complexity and it was concluded that is presents an example of a low-dimensional chaotic process [14,15]. Its properties are therefore qualitatively similar to the artificial chaotic signal studied above. To be consistent with the previous analysis, the EEG signal was normalized (average zero, variance equal to 1, using an episode free of artifacts and seizure activity as a reference). During the epileptic event there is a sudden rearrangement of the frequency composition from the broadband "normal" electric activity to a comparatively regular sequence of so-called spike-and-wave complexes (Fig. 5a). The Fourier spectrum in this section shows a dominant frequency on a basal broadband distribution as is typical not only for this type of seizures [14] but also for chaotic systems like the Rössler equation [16]. The nonstationarity of EEG in general does not allow the assignment of the label "chaotic attractor" to this episode (as distinguished for example from a noise-contaminated limit cycle). However, the signal has a waveform that is compatible with either a homoclinic [14] or a heteroclinic [15] situation both of which are generic sources of chaotic solutions [17]. In contrast to the X and Y variables from the Rössler system, however, its two time scales (the slow wave and the fast spike) give rise to a non-harmonic waveform.

The coupling is chosen such that the broadband "normal" EEG signal does not induce any excitation waves. This way the normal activity is ignored and only the seizure activity with its dominant frequencies and large amplitudes is detected. Fig. 5b is the pattern of induced waves during the time where the seizure occurs in the signal. The pattern displays three bands. The two main bands are centered initially at 3.4 Hz and 5.0 Hz, respectively. The frequency band at about 3 Hz corresponds to the dominant frequency of the spike-and-wave complexes. Clearly, the higher frequency band is not a harmonic of the lower frequency band but a second, independent frequency. Both bands undergo a shift in frequency to final values of 2.6 Hz and 4.4 Hz, respectively. This reflects the well-documented frequency slowing during absence seizures [18]. The thin third band at about 1.4 Hz starts approximately at second 192 and lasts until second 207. It is a subharmonic of the main frequency, which in this part of the seizure is centered at 2.8 Hz. The co-appearance of these two bands is reminiscent of the induction pattern for the Rössler chaos (Fig. 4). Thus, while the fine texture of the two broad bands does not permit to distinguish between noisy "limit cycle" and "chaos" as (hypothetic asymptotic) behaviors, the appearance of a subharmonic indicates a degree of complexity typical of a chaotic attractor.

The system eq. (1) in its "tuned" form exploits the properties of spatio-temporal excitable chaos to extract information from analog signals and represent them as temporally evolving excitation patterns. Some of its features are: 1) A minimum number of recurrent excitations are required to be detectable. This successfully suppresses uncorrelated noise components in the signal. 2) The sharp-edged filter characteristic (Fig. 2) permits induction of waves only in a certain frequency band. This allows a separation of contributions with distinct frequencies in the signal and estimation of their individual frequencies even in the presence of high levels of noise. 3) Onset, offset and continuous shifts in frequency of the band are features that allow almost instantaneous detection and characterization of *transient* events (see Fig. 5). 4) In contrast to Fourier spectra, wavelets, and methods derived from nonlinear dynamics (e.g. the estimation of the correlation dimension or of the spectrum of Lyapunov characteristic exponents) no pre-determined finite window of data (with assumed stationarity) is required for evaluation. The system identifies "interesting" episodes in nonstationary recordings. Its temporal resolution makes the method attractive for the analysis of time series where nonstationarity is not only unavoidable but also crucial for the understanding, e.g. EEG recordings or, in the context mentioned in the introduction, the sound-generated vibrations in the inner ear. 5) Finally, the combination of data normalization with a pre-set value of coupling constant allows *automatic* detection and characterization of specific events (e.g. sounds with characteristic overtones and undertones [2]).

The fact that the unperturbed system is excitable allows to not only detect events and changes of events in the perturbation but also - as soon as they disappear - to automatically destroy the induced coherence and to reset the basal dynamics. The method thus differs from non-resetting artificial neural networks and has dynamics closer to those of sensory neurons which are supposed to give *"a sort of running commentary"* [3] to an external signal's temporal variations. This recommends the present system for further investigation with respect to information extraction.

Work was supported by CONACyT, Mexico (project no. 40885-F). We thank U. Stephani and H. Muhle, Clinic for Neuropediatry, University of Kiel, Germany, for providing the data set in Fig. 5; and A. Galka, Institute of Experimental and Applied Physics, University of Kiel, Germany, for preprocessing of the data. We acknowledge discussions with S. Sahle and A. Galka. A data sonification of the results in Fig. 5 is available from the authors.

# Figure Captions

Fig. 1: Bifurcation diagram of eq. (1) with additive sinusoidal forcing of $X_1$ in the form *pert*=sin($\omega$t). Plotted are maxima of variables $X_1$ (a) and $X_{30}$ (b) as a function of forcing frequency $\omega$. Parameters: *a*=0.325, $m_2 = 20$, $m_3 = 23$, $k_1 = 0.8$, $k_a = 0.81$, $\varepsilon$=1.0, $D_X = 0.5$, *A*=80.

Fig. 2: Scans in parameter space of eq. (1) with *pert*=sin($\omega$t)+$\eta$, where $\eta$ denotes white noise (zero mean; variance 1). Grey-coding of number of suprathreshold maxima in variable $X_{30}$ in response to 30 sine wave cycles in the perturbation. Average of 30 runs for each point. a) Parameter plane $\varepsilon/\omega$; *A*=40. b) Parameter plane *A*/$\omega$; $\varepsilon$=1.0. Other parameters as in Fig. 1.

Fig. 3: Mean response time of variable $X_{30}$ to the onset of perturbation in variable $X_1$ as a function of $\omega$; a) expressed in seconds; b) expressed as number of perturbation cycles. For each perturbation frequency $\omega$, parameter $\varepsilon$ is calculated as $\varepsilon=\omega/2.6$ according to the calibration in Fig. 2. Plotted are averages of 30 runs at each point. Other parameters as in Fig. 1.

Fig. 4: Temporal evolution of excitations of a set of equations (1) with different parameters $\varepsilon$ perturbed by the time series of variable X from the chaotic Rössler equation. Plotted are occurrences of suprathreshold oscillations in variable $X_{30}$ as vertical bars (average of 30 runs). The time axis is corrected with the average delay in Fig. 3b. *A*=80, other parameters as in Fig. 1.

Fig. 5: a) EEG signal (electrode F4) with epileptic activity. The seizure lasts from 183-211 seconds. b) Excitations of a set of equations (1) ) with different parameters $\varepsilon$ perturbed by the EEG signal in a). Plot as in Fig. 4. *A*=120, other parameters as in Fig. 1.

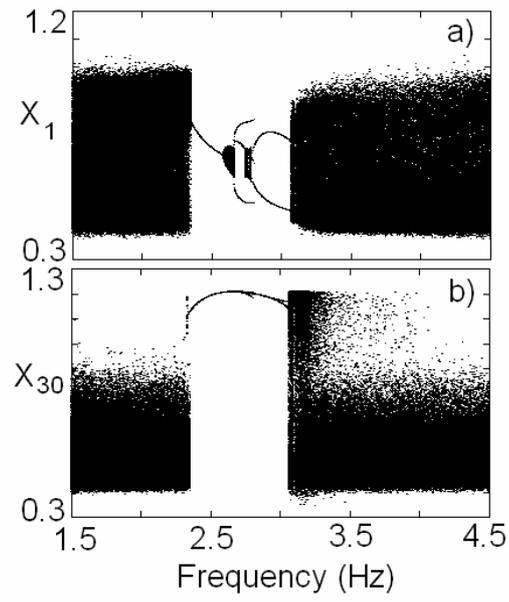

Baier and Muller, Fig. 1

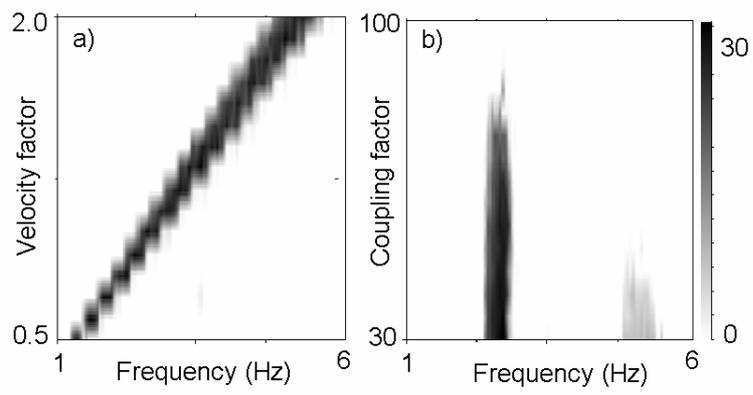

Baier and Muller, Fig. 2

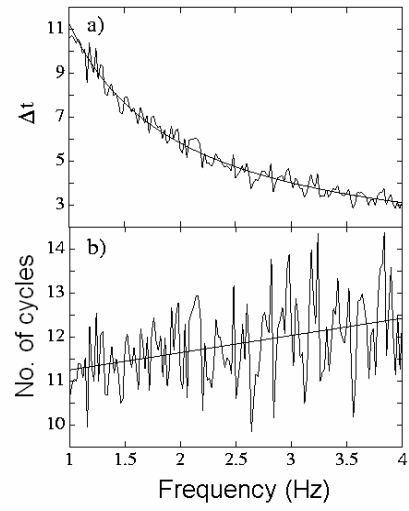

Baier and Muller, Fig. 3

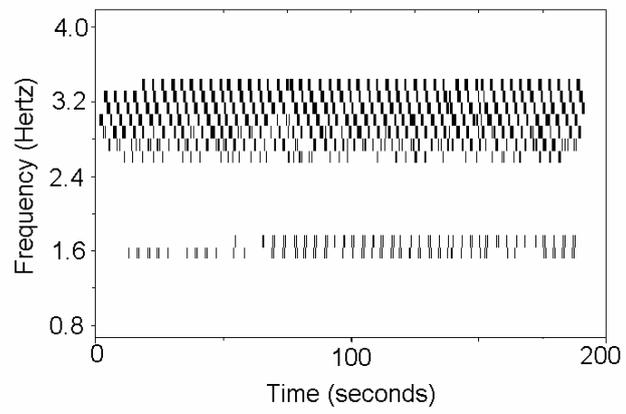

Baier and Muller, Fig. 4

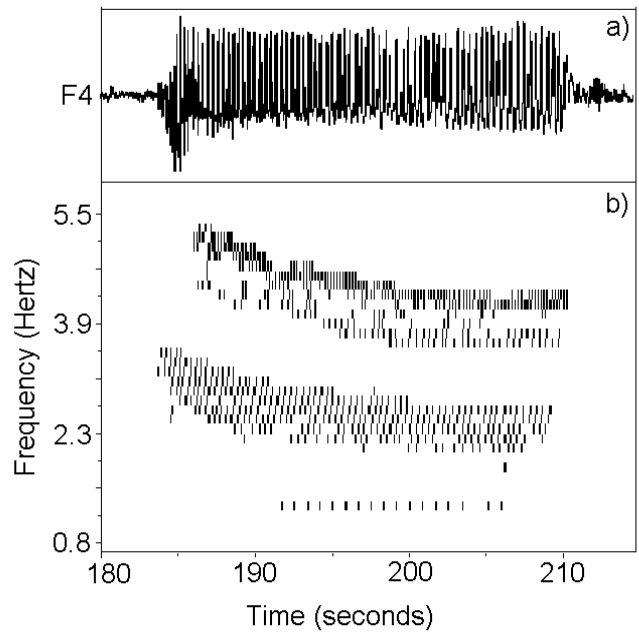

Baier and Muller, Fig. 5